\documentclass[%
 reprint,
superscriptaddress,
 amsmath,amssymb,
 aps,
 pra,
]{revtex4-2}

\usepackage{graphicx}% Include figure files
\usepackage{dcolumn}% Align table columns on decimal point
\usepackage{bm}% bold math
\usepackage{xcolor}
\definecolor{darkred}{rgb}{0.55,0,0}

\usepackage[normalem]{ulem}

\begin{document}

\preprint{APS/123-QED}

\title{Floquet framework for driven polar quantum systems}% Force line breaks with \\

\author{Viktor Novi\v{c}enko}
\email{viktor.novicenko@tfai.vu.lt}
\homepage{http://www.itpa.lt/~novicenko/}
\affiliation{Institute of Theoretical Physics and Astronomy, Vilnius University, Saul\.{e}tekio ave.~3, LT-10257 Vilnius, Lithuania}
 
\author{Piotr G\l{}adysz}
\email{glad@umk.pl}
\affiliation{Institute of Physics, Faculty of Physics, Astronomy and Informatics, Nicolaus Copernicus University in Toru\'{n}, ul. Grudziadzka 5, 87-100 Toru\'{n}, Poland}

\author{Karolina S\l{}owik}
\email{karolina@fizyka.umk.pl}
\affiliation{Institute of Physics, Faculty of Physics, Astronomy and Informatics, Nicolaus Copernicus University in Toru\'{n}, ul. Grudziadzka 5, 87-100 Toru\'{n}, Poland}%
\affiliation{Institute of Advanced Studies, Nicolaus Copernicus University in Toru\'{n}, Wile\'{n}ska 4, 87-100 Toru\'{n}, Poland}

\author{Egidijus Anisimovas}
\email{egidijus.anisimovas@ff.vu.lt}
\affiliation{Institute of Theoretical Physics and Astronomy, Vilnius University, Saul\.{e}tekio ave.~3, LT-10257 Vilnius, Lithuania}

\date{\today}% It is always \today, today,

\begin{abstract}
We present an analytical and numerical Floquet treatment of a driven polar two-level quantum system characterized by both longitudinal and transverse coupling to a periodic field. Analytically, we derive a dressed-frame effective Hamiltonian up to first order in the inverse driving frequency, incorporating the longitudinal coupling nonperturbatively. This yields closed expressions for the effective transverse coupling strength and the effective detuning, both of which are modified by the presence of the longitudinal interaction. In the nonpolar limit, these expressions recover the usual near-resonant Rabi coupling and the Bloch-Siegert shift. As a second main result, we develop a numerical flow-equation framework that yields a time-independent effective Hamiltonian across a broad range of transverse and longitudinal coupling strengths. This dual framework is relevant for a variety of platforms, including driven polar quantum systems, optical lattices, superconducting circuits, and solids subject to surface acoustic waves. 
\end{abstract}

\maketitle

%\tableofcontents

\section{\label{sec:intro}Introduction}
Driven two-level systems are among the simplest and most useful models of light-matter interaction. In the typical case, the two states have well-defined parity, and the external field couples them through a transition dipole moment. The dynamics can then be described by the Rabi model and, when the driving is not too strong and the field is near resonance, simplified by the rotating-wave approximation~\cite{Fujii2014,Fujii2017}. This gives a clear and intuitive picture: the drive induces coherent transitions between the two states, while rapidly oscillating terms are neglected.

The situation becomes richer when the system is polar, lacking inversion symmetry. In this case, the system eigenstates can possess different permanent dipole moments~\cite{Kibis2009,Macovei2015,Kryuchkyan2017,Anton2017,Bogolubov2026}. The driving field then has two effects simultaneously. It can induce transitions between the states through the usual transverse dipole matrix element, but it can also modulate the energy splitting between them through a longitudinal coupling. In other words, the transition frequency itself becomes periodically driven. This longitudinal part of the interaction is absent in inversion-symmetric systems and constitutes the key feature of the model studied in this work.

Hamiltonians of this type are realized, or can be engineered, in several physical platforms. One example is provided by driven optical lattices, where two selected states can be coupled by an external modulation while their energy separation is also periodically shifted. The parameters of this platform are used later in our numerical examples, which are based on the optical-lattice experiment of Ref.~\cite{Schweizer2019}. Similar physics arises in superconducting quantum circuits, where artificial atoms interact with microwave fields or external drives through both transverse and longitudinal couplings, and where strong- and ultrastrong-coupling regimes are experimentally accessible \cite{Niemczyk2010,Yoshihara2017,FornDiaz2019,Kockum2019}. A related route is offered by acoustic quantum systems \cite{Gustafsson2014}, where surface acoustic waves propagating through a solid act as coherent periodic perturbations coupled to electronic, spin, or superconducting degrees of freedom.

Recent investigations \cite{Burgess2023,Gladysz2025}
%, \ea{[add more]} 
have shown that driven polar systems can exhibit qualitatively new optical signatures. Permanent dipoles can strongly modify the fluorescent emission of optically dressed systems, suppressing it in selected regimes and leading to spontaneous absorption in others. 
%\ea{[Pepe]}. 
In fluorescence emission spectra, additional peaks are found compared with the traditional nonpolar case \cite{Gladysz2025}. These spectra include, in particular, a low-frequency coherent emission channel related to the permanent-dipole part of the interaction \cite{Kibis2009,Gladysz2020}.
%\ea{[Scala]}. 
Ref.~\cite{Gladysz2025} describes the driven dynamics well when the parameters remain below the ultrastrong longitudinal-coupling regime. However, ultrastrong longitudinal coupling is precisely the regime where the most characteristic spectral features are expected, and where the approximate treatment becomes less reliable.

Starting from the 19th-century work~\cite{floquet1883} on linear homogeneous differential equations with periodic coefficients, Floquet theory provides a natural framework for describing the evolution of a state vector governed by a linear differential equation with a time-periodic operator. The main idea is to perform a transformation to a Floquet frame that renders the time-periodic operator time-independent. In the case of a quantum system, the periodic operator is anti-Hermitian (the Hamiltonian divided by $i\hbar$), so in the Floquet frame it yields a static anti-Hermitian operator with a purely imaginary spectrum, known as quasienergies divided by $i\hbar$. Such a static operator (after multiplication by $i\hbar$) is known as a Floquet effective Hamiltonian. It allows one to separate the evolution into two parts: slow, long-time dynamics governed by the Floquet effective Hamiltonian, and a fast periodic micromotion due to the Floquet frame. Such a separation is most transparent when working in so-called Sambe extended space~\cite{Sambe1973}. Importantly, in the literature, the term ``Floquet Hamiltonian'' often refers to a different operator, albeit one with the same spectrum, which describes the evolution over a time interval that is an integer multiple of the driving period. For such an operator, we use the name Floquet stroboscopic Hamiltonian to emphasize its nature. In contrast to the effective Hamiltonian, the stroboscopic Hamiltonian depends on the initial phase and incorporates both the effective Hamiltonian and the periodic micromotion operator. We refer the reader to Refs.~\cite{Bukov2015,Eckardt2015} where both operators are discussed, and to Ref.~\cite{Mikami2016PRB} where the effective Hamiltonian is referred to as the van Vleck Hamiltonian and the stroboscopic as the Floquet--Magnus Hamiltonian.

Typically, finding the Floquet effective Hamiltonian is as difficult as solving the exact Schr\"{o}dinger equation; therefore, an approximate solution is required. One common choice is the high-frequency approach, in which the effective Hamiltonian can be systematically constructed as an expansion in inverse powers of the driving frequency~\cite{Rahav2003,Guerin2003,Goldman2014,Novicenko2017,Mikami2016PRB}. However, for the polar system considered here, a direct high-frequency expansion of the laboratory-frame Hamiltonian is not justified since the longitudinal coupling is taken to be comparable to the driving frequency.

In this work, we develop a Floquet framework for a driven polar two-level system in the regime of ultrastrong longitudinal coupling. We transform the laboratory-frame Hamiltonian to a dressed frame, where the longitudinal coupling is retained nonperturbatively while the transverse coupling serves as the small parameter for a perturbative expansion.

We then derive an effective Floquet Hamiltonian up to first order in the inverse driving frequency, together with the corresponding first-order micromotion operator. We obtain expressions for an effective transverse coupling and effective detuning, both of which are modified in the presence of the longitudinal coupling. In the nonpolar limit, our theory recovers the usual near-resonant Rabi coupling and the Bloch--Siegert shift.

Finally, we benchmark our analytical results against the numerical effective Hamiltonian obtained from flow equations and against direct integration of the laboratory-frame Schr\"{o}dinger equation.

\section{\label{sec:model}Model: a driven polar two-level system}

We consider a driven two-level quantum system with ground and excited states $\left|g\right\rangle$ and $\left|e\right\rangle$, 
separated by the transition frequency $\omega_{eg}$. The system is driven by a classical monochromatic electric field 
$\mathbf{E}(t)=\mathbf{E}_0 \cos(\omega t)$, with driving frequency $\omega$ close to resonance with the transition, $\omega \simeq \omega_{eg}$. 
The light-matter interaction is described in the electric-dipole approximation by the dipole moment operator 
$\hat{\mathbf{d}}=\sum_{i,j \in \left\{g,e \right\}} \mathbf{d}_{i,j} \left|i \right\rangle \left\langle j\right|$, where 
$\mathbf{d}_{i,j}=q\left\langle i\right| \hat{\mathbf{r}}\left|j \right\rangle$ and $q$ is the electron charge. 

In systems with inversion symmetry, the diagonal dipole matrix elements vanish, $\mathbf{d}_{g,g}=\mathbf{d}_{e,e}=\mathbf{0}$, and the drive 
couples the two states only through the transition dipole moment $\mathbf{d}_{e,g}$. Equivalently, the undriven Hamiltonian is invariant 
under the inversion through the origin $\mathbf{r}\mapsto-\mathbf{r}$, or $[\hat{H}_0,\hat{\Pi}]=0$, where $\hat{\Pi}$ is the parity operator. 
$\hat{H}_0$ eigenstates can therefore be chosen to transform according to 
the irreducible representations of the two-element group:
%\ea{$C_i=\{e,i\}$}: 
the trivial
representation corresponding to even states, and the sign representation corresponding to odd states. Since the dipole operator is odd under inversion,
its diagonal matrix elements vanish by parity.

Here, we consider instead a polar quantum system, for which inversion symmetry is broken.
As a result, the two states may carry different permanent dipole moments,
$\mathbf{d}_{e,e}\neq \mathbf{d}_{g,g}$, giving rise to a longitudinal coupling 
to the driving field. Such a coupling is absent in the standard inversion-symmetric
Rabi model and is the central feature of the driven systems considered below.

The resulting laboratory-frame Hamiltonian takes the form
\begin{equation}
\hat{H}_{\mathrm{lab}}(\omega t)=\frac{1}{2}[\hbar \omega_{eg} +\hbar g_z \cos(\omega t)]\hat{\sigma}_z+\hbar g_x \cos(\omega t)\hat{\sigma}_x,
\label{eq:main}
\end{equation}
where $\hbar g_z=\mathbf{E}_0\cdot \left(\mathbf{d}_{e,e}-\mathbf{d}_{g,g}\right)$ is the longitudinal coupling strength, and 
$\hbar g_x=\mathbf{E}_0\cdot\mathbf{d}_{e,g}$ 
represents the standard transverse coupling strength. Note the different physical impact of the two coupling terms: While the transverse coupling induces transitions between the eigenstates, the longitudinal one results in rapid oscillations of the diagonal Hamiltonian terms, which can be interpreted as modulation of the transition frequency.

In the following, we focus on the near-resonant regime
$\omega_{eg}-\omega=\mathcal{O}(1)$, while allowing the longitudinal
coupling to be comparable to the driving frequency, $g_z=\mathcal{O}(\omega)$.
The transverse coupling, by contrast, is kept perturbative,
$g_x=\mathcal{O}(1)$. This regime cannot be treated by a direct
high-frequency expansion of Eq.~(\ref{eq:main}), because both $\omega_{eg}$ and
$g_z$ are comparable with the driving frequency. The analytical Floquet construction
developed below therefore starts from a dressed rotating frame in which the
longitudinal modulation is absorbed nonperturbatively.

%\textcolor{blue}{Goals not clear. Describe here or in Introduction. \\
%We need here a description of approaches to be compared: Floquet analysis at different orders vs numerical solutions. Complicated Hamiltonian with rapid time dependence solved analytically. Effective Hamiltonian derived of simple structure, with time-independent operator terms. Difficult things made easy.\\
%Throughout the paper some guidance on what is done why.}

\section{\label{sec:floquet}Analytical effective Floquet Hamiltonian for longitudinally dressed systems}
Van Vleck high-frequency expansion provides a systematic way of constructing a time-independent Floquet effective Hamiltonian and the associated micromotion
operator as a series in inverse powers of the driving frequency  \cite{Eckardt2015,Mikami2016PRB}. The expansion assumes a separation of time
scales: the driving period $T=2\pi/\omega$ should be short compared with the
time scale of the effective dynamics, and the Fourier components of the
Hamiltonian used as the starting point should remain bounded in the
high-frequency limit~\cite{Novicenko2017,Novicenko2022}. In practice,
this means that all Hamiltonian parameters should be small compared to $\omega$.

The laboratory-frame Hamiltonian in Eq.~(\ref{eq:main}) does not satisfy this
condition in the parameter regime considered here. Since the drive is close to
resonance with the two-level transition, the transition frequency is comparable
to the driving frequency. Moreover, in a polar system the longitudinal
modulation can also be strong, with $g_z$ comparable to $\omega$. Thus, large
energy scales appear explicitly in the Fourier amplitudes of
$\hat{H}_{\mathrm{lab}}(\omega t)$.

To restore a useful high-frequency hierarchy, we first transform the Hamiltonian
to a frame in which these large contributions are absorbed. The near-resonant
transition frequency is treated by the usual rotating-frame transformation~\cite{Fujii2014,Fujii2017}, while the longitudinal modulation is incorporated
through an additional dressing transformation~\cite{Kibis2009}. The resulting Hamiltonian retains the ratio $g_z/\omega$ nonperturbatively, while the remaining time-dependent Fourier amplitudes are proportional to the
transverse coupling $g_x$. The analytical expansion is therefore accurate when
$g_x/\omega$ is small, even if $g_z/\omega$ is not.

\subsection{\label{subsec:frame_transformation}Frame transformation}
We first write the laboratory-frame Hamiltonian as a Fourier series,
\begin{equation}
\hat{H}_{\mathrm{lab}}(\omega t)
=
\sum_{m=-\infty}^{+\infty}
\hat{H}_{\mathrm{lab}}^{(m)}
e^{im\omega t}.
\label{eq:fourexp}
\end{equation}
Rather than applying the high-frequency expansion directly to these Fourier
components, we introduce two unitary transformations.
The first one moves the system to a frame that rotates sinusoidally about the $z$ axis~\cite{Kibis2009}
\begin{equation}
\hat{U}_{1}(\omega t)
=
\exp\left[
    \frac{i g_z}{2\omega}
    \sin(\omega t)
    \hat{\sigma}_z
\right].
\label{eq:U_long}
\end{equation}
This transformation removes the explicitly time-dependent diagonal term
proportional to $g_z\cos(\omega t)$ from the Hamiltonian and incorporates the
longitudinal modulation into the phase of the transverse coupling~\cite{Kibis2009,Gladysz2025}. The second transformation
is the standard passage to a frame rotating at the driving frequency
\begin{equation}
\hat{U}_{\mathrm{rot}}(\omega t)=\exp\left[\frac{i }{2}\omega t \hat{\sigma}_z\right].
\label{eq:U1}
\end{equation}
The transformed Hamiltonian in the interaction picture $\hat{U}(\omega t)=\hat{U}_{\mathrm{rot}}(\omega t) \hat{U}_{1}(\omega t)$ reads
\begin{equation}
\begin{aligned}
&\hat{H}(\omega t) = \hat{U}(\omega t) \hat{H}_{\mathrm{lab}}(\omega t) \hat{U}^{\dagger}(\omega t)+i\hbar\frac{d\hat{U}(\omega t)}{dt}\hat{U}^{\dagger}(\omega t) \\
&= \frac{\hbar}{2}(\omega_{eg}-\omega)\hat{\sigma}_z+\hbar g_x \cos(\omega t) \left( \exp\left[i\frac{g_z}{\omega} \sin(\omega t) \right] e^{i\omega t} \hat{\sigma}_+ \right. \\
&+ \left.\exp\left[-i\frac{g_z}{\omega} \sin(\omega t) \right] e^{-i\omega t} \hat{\sigma}_-\right),
\end{aligned}
\label{eq:int_pic}
\end{equation}
where $\hat{\sigma}_{\pm}=(\hat{\sigma}_x \pm i \hat{\sigma}_y)/2$. 
In Eq.~(\ref{eq:int_pic}), large transition energy has been reduced to the detuning
$\omega_{eg}-\omega$ while the longitudinal modulation is retained
nonperturbatively in the phase factors
$\exp[\pm i(g_z/\omega)\sin(\omega t)]$. The remaining time-dependent
amplitudes are proportional to the relatively small transverse coupling $g_x$. This transformed
Hamiltonian is therefore the appropriate starting point for the
Floquet-van Vleck expansion in the regime considered here.

\subsection{\label{subsec:expansion}Floquet effective Hamiltonian and micromotion through high-frequency expansion}
We now expand the transformed Hamiltonian (\ref{eq:int_pic}) into Fourier series (\ref{eq:fourexp}).
The phase factors generated by the longitudinal dressing are expanded using the
Jacobi-Anger identity $e^{iz\sin\theta}=\sum_{n=-\infty}^{+\infty}J_n(z)e^{in\theta}$, 
where $J_n$ denotes the Bessel function of the first kind. 
This gives the Fourier components
\begin{equation}
\hat{H}^{(0)}=\frac{\hbar}{2}(\omega_{eg}-\omega)\hat{\sigma}_z+\hbar g_x \frac{\omega}{g_z} J_1 \left( \frac{g_z}{\omega} \right) \hat{\sigma}_x,
\label{eq:H0}
\end{equation}
and
\begin{equation}
\begin{aligned}
\hat{H}^{(m\neq 0)} =& \hbar g_x\frac{\omega}{g_z} \left[ (m-1)J_{m-1}\left( \frac{g_z}{\omega}\right) \hat{\sigma}_+ \right. \\
&+ \left.(m+1)J_{m+1}\left( \frac{g_z}{\omega}\right) (-1)^{m} \hat{\sigma}_-\right].
\end{aligned}
\label{eq:Hm}
\end{equation}
The zeroth harmonic already contains a nontrivial renormalization of the
transverse coupling,
$g_x\to (g_x\omega/g_z)J_1(g_z/\omega)$. This term gives the leading effective
Rabi coupling in the longitudinally dressed frame. Higher Fourier components
generate corrections to this leading dynamics through the van Vleck expansion.
Note that in the absence of longitudinal coupling, the zeroth harmonic reduces to
the usual transverse coupling.

The Fourier components in Eqs.~(\ref{eq:H0}) and~(\ref{eq:Hm}) remain bounded
in the high-frequency limit. We can therefore apply the inverse-frequency
expansion to obtain both the Floquet effective Hamiltonian and the micromotion
operator~\cite{Novicenko2017,Novicenko2022}.

%\sout{
Note that the term "Floquet Hamiltonian" can refer to many different, yet closely related, operators. In the context of high-frequency expansion, one usually distinguishes between stroboscopic and effective Floquet Hamiltonians~\cite{Bukov2015}. The former can be obtained
from the Magnus expansion and describes the evolution over integer numbers of
driving periods and generally depends on the choice of initial phase of the drive. The
van Vleck effective Hamiltonian used here is phase-independent~\cite{Rahav2003};
%}
%\ea{As discussed in the Introduction, we use the phase-independent van Vleck effective Hamiltonian~\cite{Rahav2003}.}
the full evolution is recovered by combining the effective dynamics with
micromotion operators evaluated at the initial and final times.
This evolution from time $t_0$ to $t_1$ can be then factorized as
\begin{equation}
\hat{U}_{\mathrm{evol}}(t_1,t_0)=\hat{U}_{\mathrm{mic}}(\omega t_1) \hat{U}_{\mathrm{eff}}(t_1,t_0) \hat{U}^{\dagger}_{\mathrm{mic}}(\omega t_0),
\label{eq:evol}
\end{equation}
where the effective evolution is generated by the static effective Floquet Hamiltonian 
\mbox{$\hat{U}_{\mathrm{eff}}(t_1,t_0)=\exp[(i\hbar)^{-1}\hat{H}_{\mathrm{eff}}(t_1-t_0)]$}, 
and the micromotion operator is written as 
$\hat{U}_{\mathrm{mic}}(\omega t)=\exp[-i \hat{S}_{\mathrm{mic}}(\omega t)]$.
Here, $\hat{S}_{\mathrm{mic}}(\omega t)$ is a Hermitian, periodic operator
$\hat{S}_{\mathrm{mic}}(\omega t +2\pi)=\hat{S}_{\mathrm{mic}}(\omega t)$, 
with zero average over one driving period $\int_0^{2\pi} \hat{S}_{\mathrm{mic}}(\theta) d\theta =\hat{0}$.
The high-frequency expansion of both $\hat{H}_{\mathrm{eff}}$ and $\hat{S}_{\mathrm{mic}}(\omega t)$ can be obtained within the van Vleck construction by block diagonalization in the extended, or Sambe, space~\cite{Sambe1973,Eckardt2015}.

To zeroth order, the effective Hamiltonian is the zero Fourier component of the transformed Hamiltonian $\hat{H}_{\mathrm{eff}(0)}=\hat{H}^{(0)}$, presented in Eq.~(\ref{eq:H0}).
At this order, the micromotion vanishes $\hat{S}_{\mathrm{mic}(0)}=\hat{0}$. 
This leading-order result is consistent with the effective Hamiltonian obtained
in Ref.~\cite{Gladysz2025}, recovering the effective transverse
coupling obtained there through a similar procedure. The corresponding transition-frequency
renormalization contributes only at higher order in the present inverse-frequency expansion.

The first-order van Vleck correction is determined by the commutators of
opposite Fourier harmonics
\begin{equation}
\begin{aligned}
\hat{H}_{\mathrm{eff}(1)} &= \frac{1}{\hbar \omega} \sum_{m=1}^{+\infty} \frac{[\hat{H}^{(m)},\hat{H}^{(-m)}]}{m}\\
=& \hat{\sigma}_z\frac{4(\hbar g_x)^2}{\hbar\omega} \left( \frac{ \omega}{g_z} \right)^2 \sum_{m=1}^{+\infty} \left[J_m^\prime\left( \frac{g_z}{\omega} \right)-\frac{\omega}{g_z} J_m\left( \frac{g_z}{\omega} \right) \right] \\
& \times \left[m^2 \frac{\omega}{g_z} J_m\left( \frac{g_z}{\omega} \right) - J_m^\prime\left( \frac{g_z}{\omega} \right) \right].
\end{aligned}
\label{eq:eff1}
\end{equation}
In the last expression, the infinite sum can be simplified using relations of the Bessel functions
\begin{equation}
\begin{aligned}
\hat{H}_{\mathrm{eff}(1)}=& \hat{\sigma}_z \frac{(\hbar g_x)^2}{\hbar\omega}\left(\frac{\omega}{g_z}\right)^2\\
& \times \left[2 J_1^2\left( \frac{g_z}{\omega} \right) +2 \frac{\omega}{g_z}J_0\left( \frac{g_z}{\omega} \right) J_1\left( \frac{g_z}{\omega} \right)-1 \right].
\end{aligned}
\label{eq:eff1_sim}
\end{equation}
This correction is diagonal in the two-level basis and therefore gives the
leading correction to the effective detuning. 
In the nonpolar limit,
$g_z\rightarrow 0$, $\hat{H}_{\mathrm{eff}(1)}$ reduces to the Bloch-Siegert shift
\begin{equation}
\lim_{g_z \rightarrow 0} \hat{H}_{\mathrm{eff}(1)}=\hat{\sigma}_z\frac{(\hbar g_x)^2}{8\hbar\omega}.
\label{eq:eff1_lim}
\end{equation}
The first-order correction to the micromotion operator reads
\begin{equation}
\begin{aligned}
\hat{S}_{\mathrm{mic}(1)}(\omega t) =& \frac{1}{i\hbar \omega} \sum_{m\neq 0} \frac{\hat{H}^{(m)}}{m}e^{im\omega t}\\
=& \frac{g_x}{g_z} \sum_{m=1}^{+\infty} \left\lbrace\hat{\sigma}_x \sin(m\omega t) \left[ X_m + (-1)^m Y_m  \right] \right.\\
&+\left. \hat{\sigma}_y \cos(m\omega t) \left[ X_m + (-1)^{m+1} Y_m  \right] \right\rbrace.
\end{aligned}
\label{eq:mic1}
\end{equation}
where $X_m=\frac{m-1}{m}J_{m-1}\left( \frac{g_z}{\omega} \right)$ and $Y_m=\frac{m+1}{m}J_{m+1}\left( \frac{g_z}{\omega} \right)$.

Now we can recover the full evolution operator in the laboratory frame~(\ref{eq:main}) as
\begin{equation}
\begin{aligned}
\hat{U}_{\mathrm{lab}}(t_1,t_0) =& \hat{U}_1^\dagger(\omega t_1)\hat{U}_{\mathrm{rot}}^\dagger(\omega t_1) e^{-i \hat{S}_{\mathrm{mic}(1)}(\omega t_1)} \\
& \times \exp\left[\frac{\hat{H}_{\mathrm{eff}(0)}+\hat{H}_{\mathrm{eff}(1)}}{i\hbar}(t_1-t_0)\right] \\
& \times e^{i \hat{S}_{\mathrm{mic}(1)}(\omega t_0)} \hat{U}_{\mathrm{rot}}(\omega t_0) \hat{U}_1(\omega t_0).
\end{aligned}
\label{eq:ev_all}
\end{equation}

Equations~(\ref{eq:H0}),~(\ref{eq:eff1_sim}), and~(\ref{eq:mic1}) provide the analytical Floquet description of the driven polar two-level system in the transformed frame. They yield a time-independent effective Hamiltonian up to first order in the high-frequency expansion, with the longitudinal coupling retained nonperturbatively, and the corresponding first-order micromotion. Equation~(\ref{eq:ev_all}) then reconstructs the laboratory-frame evolution. In the following, we develop a numerical flow-equation construction of the effective Hamiltonian and use it to benchmark the analytical results and identify their domain of validity.

\section{\label{sec:numerical_hamiltonian}Numerical construction of the effective Hamiltonian}
We now construct numerically the time-independent effective Hamiltonian
associated with the transformed Hamiltonian in Eq.~(\ref{eq:int_pic}). For the two-level system considered here, this Hamiltonian can be written, up to an
irrelevant term proportional to the identity, as
\begin{equation}
\hat{H}_{\mathrm{eff}} = \hbar\sum_{j=x,y,z}g^{\mathrm{eff}}_j\hat{\sigma}_j .
\label{eq:Heff_components}
\end{equation}
The coefficients $g^{\mathrm{eff}}_j$ describe the effective interaction in the
longitudinally dressed rotating frame. They can be obtained analytically from
the high-frequency expansion, as shown in Sec.~\ref{sec:floquet}, or computed
numerically from a truncated set of Fourier harmonics.

The numerical construction is useful for two reasons. First, it provides an
independent check of the analytical expressions derived above. Second, it gives
access to a description beyond a fixed-order inverse-frequency expansion. 
To the best of the authors' knowledge, this is the first attempt to derive a methodology for numerically calculating the phase-independent effective Hamiltonian for a finite driving frequency.

\subsection{\label{subsec:flow_method}Flow-equation method}
The numerical calculation of $\hat{H}_{\mathrm{eff}}$ for the finite frequency $\omega$ is a very non-trivial task: Even if full evolution data can be easily obtained by direct integration of the Schr\"{o}dinger equation, it is not clear how to factorize Eq.~(\ref{eq:evol}) and split the dynamics due to micromotion from the dynamics due to effective evolution. By definition~\cite{Rahav2003,Goldman2014}, the micromotion operator $\hat{S}_{\mathrm{mic}}(\omega t)$ has zero average over one period, but that does not imply any useful information about the unitary micromotion operator $\hat{U}_{\mathrm{mic}}(\omega t)$. 

The idea presented below is based on the work~\cite{Novicenko2022} and uses the numerical integration of flow equations.
If the Hilbert space of the system is finite% (in our case, it is $\mathbb{C}^2$)
, and if the periodic Hamiltonian has a limited number of non-zero harmonics in the Fourier expansion, one can integrate the flow equations derived in Ref.~\cite{Novicenko2022} until all harmonics $\hat{H}^{(m)}(s)$ settle to fixed values. Technically speaking, for the Hamiltonian~(\ref{eq:int_pic}), all harmonics are non-zero, but one can see from Eq.~(\ref{eq:Hm}) that for an $m$ high enough, say $m_0$, the harmonics $\hat{H}^{(m\geq m_0)}$ become negligible due to the high order of the Bessel functions. 

As shown in Ref.~\cite{Novicenko2022,Verdeny2013}, the effective Hamiltonian can be obtained by a
continuous block diagonalization of the Hamiltonian represented in the extended, or Sambe, space. 
The flow parameter $s$ labels the progress of this block diagonalization. At $s=0$, the Fourier components are those of the dressed-frame Hamiltonian, Eqs.~(\ref{eq:H0}) and~(\ref{eq:Hm}). In the limit $s\rightarrow+\infty$, the nonzero harmonics are suppressed $\hat{H}^{(m\neq0)}(s\rightarrow+\infty)=\hat{0}$, while the zero harmonic converges to the effective Hamiltonian $\hat{H}^{(0)}(s\rightarrow+\infty)=\hat{H}_{\mathrm{eff}}$.

The flow equations used to perform the diagonalization can be derived using different flow generators. For example, Wegner-type generator used in Ref.~\cite{Verdeny2013} gives the flow equations that couple the $m$th Fourier harmonic with all other Fourier harmonics. In contrast, Ref.~\cite{Novicenko2022} provides the so-called Toda-type generator, which gives flow equations where the $m$th harmonic is coupled only with the harmonics lower or equal to $|m_0|$. It means that the harmonic truncation up to $m_0$ produces correct diagonalization if $m_0$ is high enough. The flow equations read, see Ref.~\cite{Novicenko2022} Eqs.~(53) or Eqs.~(E7)-(E8):
\begin{equation}
\begin{aligned}
\frac{d\hat{H}^{(0)}(s)}{ds} =&  \frac{2}{\hbar\omega} \sum_{m=1}^{m_0} [\hat{H}^{(m)}(s),\hat{H}^{(m)\dagger}(s)], \\
\frac{d\hat{H}^{(m \neq 0)}(s)}{ds} =& -m \hat{H}^{(m)}(s) + \frac{[\hat{H}^{(m)}(s),\hat{H}^{(0)}(s)]}{\hbar \omega} \\
 &+\frac{2}{\hbar \omega} \sum_{l=1}^{m_0-m} [\hat{H}^{(m+l)}(s),\hat{H}^{(l)\dagger}(s)],
\end{aligned}
\label{eq:flow}
\end{equation}
where $m_0$ is the largest harmonic retained in the truncated Fourier expansion.
Only the harmonics with $m\geq 0$ need to be integrated explicitly, since Hermiticity gives
$\hat{H}^{(-m)}(s)=\hat{H}^{(m)\dagger}(s)$.

\begin{figure}[h!]
    \centering
    \includegraphics[width=0.99\columnwidth]{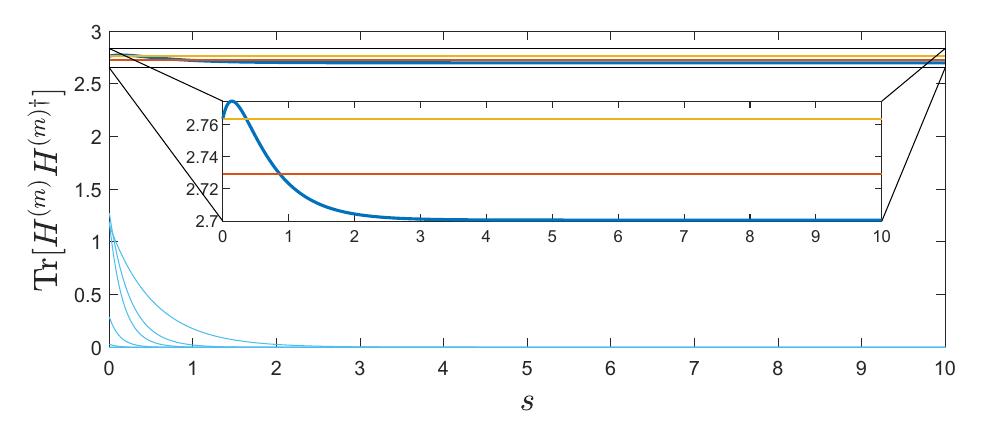}
    \caption{Harmonic weights obtained from the numerical integration of the flow equations
(blue lines) and from the analytical effective Hamiltonians: the zeroth-order Hamiltonian, Eq.~(\ref{eq:H0}) (yellow line), and the zeroth-plus-first-order Hamiltonian, Eqs.~(\ref{eq:H0}) and~(\ref{eq:eff1_sim}) (red line). Light blue lines correspond to nonzero harmonics, $m>0$, while the dark blue line corresponds to the zero harmonic. The inset shows the convergence of the
zero-harmonic weight. Including the first-order correction brings the analytical result closer to the numerical effective Hamiltonian. The parameters are $\omega=4.122 \times 2\pi$ kHz, $\omega_{eg}-\omega=0.2\times2\pi$ kHz, $g_z=0.8\omega$, and $g_x=0.5\times2\pi$ kHz, chosen in accordance with the atomic-lattice parameters used in the caption of Fig.~2 of Ref.~\cite{Schweizer2019}.}
    \label{fig_app_1}
\end{figure}

Figure~\ref{fig_app_1} shows the numerical integration of the flow equations
through the weights $\mathrm{Tr}\left[\hat{H}^{(m)}(s)\hat{H}^{(m)\dagger}(s)\right]$ of the Fourier harmonics.  For this integration, we take $m_0=10$ harmonics since for higher values of $m$, even at $s=0$, the weight of the harmonic is almost zero. We find that the weights of the nonzero harmonics settle to zero, while the zero harmonic $\hat{H}^{(0)}(s)$ converges to a time-independent matrix. This limiting matrix is identified with the (numerically obtained) Floquet effective Hamiltonian in the transformed frame.
In this figure, the parameters were selected to match the experimentally realized optical lattice system described in Ref.~\cite{Schweizer2019}.

\subsection{\label{subsec:validation}Validation}
\begin{figure}[h]
    \centering
    \includegraphics[width=0.99\columnwidth]{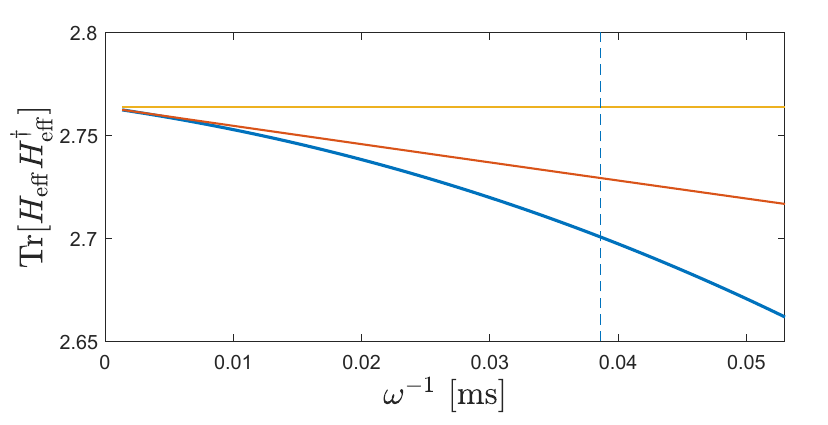}
    \caption{Effective Hamiltonian weights obtained numerically as a function of inverse driving frequency (blue line), in comparison with the analytical zeroth-order result (yellow line), and the zeroth-plus-first-order result (red line). The parameters are the same as in Fig.~\ref{fig_app_1}. The dashed vertical line marks $\omega=4.122\times2\pi$ kHz, used in Fig.~\ref{fig1}. \label{fig_app_2}}
\end{figure}

To further test the analytical expansion, we have computed the effective Hamiltonian weights
numerically for different values of the driving frequency and compared them with
the ones obtained for the analytical $\hat{H}_{\mathrm{eff}(0)}$ (Eq.~(\ref{eq:H0})) and $\hat{H}_{\mathrm{eff}(0)}+\hat{H}_{\mathrm{eff}(1)}$ (Eqs.~(\ref{eq:H0}) and (\ref{eq:eff1})). The comparison is shown in Fig.~\ref{fig_app_2}. 
The zeroth-order Hamiltonian weight (shown in yellow) does not depend on the inverse frequency and gives the limiting value at $\omega^{-1}\rightarrow 0$ for the numerical effective Hamiltonian (shown in blue).
The first-order correction gives the leading slope of the numerical result around this point. Thus, the zeroth-plus-first-order expression (red line) is tangent to the numerical curve in the high-frequency limit $\omega^{-1}\rightarrow 0$, as expected from the inverse-frequency expansion.
This confirms the correctness of the expressions~(\ref{eq:H0}) and (\ref{eq:eff1}).

%In order to check the expressions for the zero-order, Eq.~(\ref{eq:H0}), and for the first-order, Eq.~(\ref{eq:eff1}), effective Hamiltonian, in Fig.~\ref{fig_app_2} using the flow equations we calculated numerically the effective Hamiltonian for various frequencies $\omega$ and compare it with $\hat{H}_{\mathrm{eff}(0)}$ and $\hat{H}_{\mathrm{eff}(0)}+\hat{H}_{\mathrm{eff}(1)}$. As can be seen, the yellow line ($\hat{H}_{\mathrm{eff}(0)}$) does not depend on the inverse frequency and coincides with the blue line (exact $\hat{H}_{\mathrm{eff}}$) at a point $\omega^{-1}=0$, while the red line ($\hat{H}_{\mathrm{eff}(0)}+\hat{H}_{\mathrm{eff}(1)}$) is a tangential line for the blue line at the point $\omega^{-1}=0$. This confirms the correctness of the expressions~(\ref{eq:H0}) and (\ref{eq:eff1}).

%\textcolor{purple}{As an initial condition to the flow equations~(\ref{eq:flow}) we use the interactive picture Hamiltonian~(\ref{eq:int_pic}), yet it seems that the starting Hamiltonian~(\ref{eq:main}) can also be used for the integration of the flow equations. Moreover, Eq.~(\ref{eq:main}) has a definite limiting harmonic number $m_0=1$, so one can raise the question whether Eq.~(\ref{eq:int_pic}) is better than Eq.~(\ref{eq:main})? The point is that the flow equations does not necessary converge to fixed point, as it happens in Fig.~\ref{fig_app_1}. The negative feedback term, $-m\hat{H}^{(m)}(s)$, on the right hand side of Eq.~(\ref{eq:flow}) tends to minimize the weight of the harmonic $\hat{H}^{(m\neq 0)}(s)$ while the commutator term might tend to maximize such weight. Since we are interested in the parameter values $g_z$ comparable with $\omega$, in case of Eq.~(\ref{eq:main}), the commutator term might dominate over $-m\hat{H}^{(m)}(s)$, and as a consequence the convergence will not be achieved, whereas the case~(\ref{eq:int_pic}) is free from these disadvantages.}

The flow equations are initialized with the Fourier components of the transformed Hamiltonian in Eq.~(\ref{eq:int_pic}). In principle, one could attempt to apply the same procedure directly to the laboratory-frame Hamiltonian in Eq.~(\ref{eq:main}), whose Fourier expansion contains only the harmonics $m=0,\pm1$. 
However, this is not advantageous in the strong-longitudinal-coupling regime. 
The convergence of the flow results from a competition between two terms:
The negative feedback term, $-m\hat{H}^{(m)}(s)$, on the right hand side of Eq.~(\ref{eq:flow}) tends to minimize the weight of the nonzero harmonics $\hat{H}^{(m\neq 0)}(s)$, while the commutator term might tend to maximize such weight.
Since we are interested in the parameter values $\omega_{eg}$ and $g_z$ comparable with $\omega$, the commutator terms may dominate and the flow need not converge to a fixed point. The dressed-frame Hamiltonian avoids this problem by absorbing the near-resonant rotation and the longitudinal
modulation before the flow is applied.

\section{\label{sec:validity}Comparison of approaches}
Having obtained the analytical effective Hamiltonian and its numerical
flow-equation counterpart, we now compare the resulting dynamics and identify
the parameter range in which the analytical expansion remains accurate. The key
distinction is between the two couplings: the longitudinal coupling $g_z$ is
included nonperturbatively through the dressing transformation, whereas the
transverse coupling $g_x$ enters the high-frequency expansion perturbatively.
We therefore expect the analytical expressions to remain reliable for finite
$g_z/\omega$, provided that $g_x/\omega$ remains sufficiently small.

\subsection{Benchmark against direct time evolution}
\label{subsec:direct_dynamics}
\begin{figure}[h!]
    \centering
    \includegraphics[width=0.99\columnwidth]{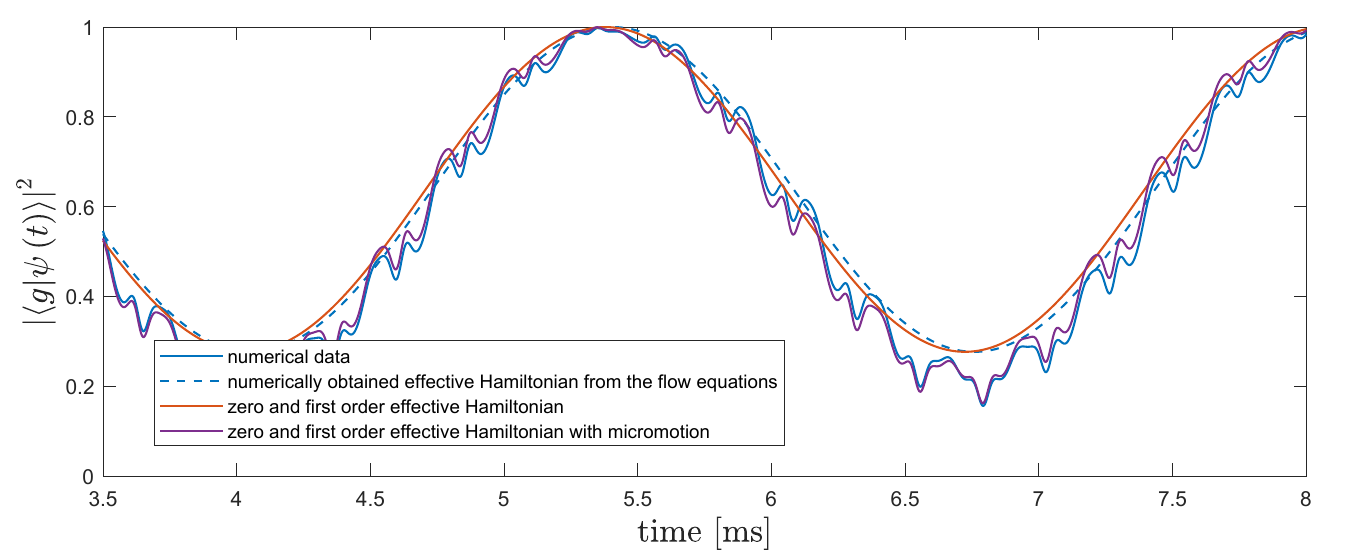}
    \caption{Comparison of the dynamics obtained from direct integration of the Schr\"odinger equation generated by the Hamiltonian in Eq.~(\ref{eq:main}) (blue solid line) with effective Hamiltonian descriptions: the Floquet effective Hamiltonian obtained from the integration of the flow equations (blue dashed line), analytically obtained Floquet effective Hamiltonian up to the first order, Eqs.~(\ref{eq:H0}) and (\ref{eq:eff1}) without (red solid line) or with micromotion included analytically up to the first order (Eq.~(\ref{eq:mic1}), purple solid line). The initial state is $\left|\psi(0)\right\rangle=\left|g\right\rangle$. Parameters as in Fig.~\ref{fig_app_1}. \label{fig1}}
\end{figure}

We compare the analytical laboratory-frame evolution operator in
Eq.~(\ref{eq:ev_all}), including first-order corrections and micromotion, with direct numerical integration of the Schr\"odinger equation generated by the laboratory-frame Hamiltonian in Eq.~(\ref{eq:main}). The comparison is shown in Fig.~\ref{fig1}. In addition to the direct
Schr\"odinger evolution, we show the dynamics generated by the numerical
flow-equation effective Hamiltonian and by the analytical effective Hamiltonian
truncated at the first order, with and without micromotion. 

This comparison separates two levels of approximation. The effective Hamiltonian 
alone (either analytical $\hat{H}_{\mathrm{eff}(0)}+\hat{H}_{\mathrm{eff}(1)}$ or numerical $\hat{H}_{\mathrm{eff}}$) captures the slow dynamics in the transformed frame, while the
micromotion factors in Eq.~(\ref{eq:ev_all}) restore the fast oscillations. 

%Since all parts of the evolution operators are obtained analytically, we can compare Eq.~(\ref{eq:ev_all}) and the numerical simulations. To do that, we integrate the full Schrodinger's equation and plot the results in Fig~\ref{fig1}. Additionally, in the same figure we numerically obtain the effective Hamiltonian $\hat{H}_{\mathrm{eff}}$ and compare the dynamics with the analytical expansion $\hat{H}_{\mathrm{eff}(0)}+\hat{H}_{\mathrm{eff}(1)}$.
%

\subsection{\label{subsec:micromotion}Micromotion reconstruction}
We next test the analytical micromotion operator in Eq.~(\ref{eq:mic1}). While the numerical calculation of the micromotion operator is a non-trivial task, we perform it with the help of a numerically obtained effective Hamiltonian $\hat{H}_{\mathrm{eff}}$. For each initial phase $\omega t\in[0,2\pi)$, we compute the stroboscopic evolution over one driving period $\hat{U}_{\mathrm{strob}}^{[t]}=\hat{U}_{\mathrm{evol}}(t+T,t)=\exp[(i\hbar)^{-1} \hat{H}_{\mathrm{strob}}^{[t]}T]$. 

Using the factorization in Eq.~(\ref{eq:evol}), this operator can be written as
\begin{equation}
\hat{U}_{\mathrm{strob}}^{[t]}=e^{-i\hat{S}_{\mathrm{mic}}(\omega t)}e^{\frac{\hat{H}_{\mathrm{eff}}}{i\hbar} T} e^{i\hat{S}_{\mathrm{mic}}(\omega t)}.
\label{eq:strob}
\end{equation}

Since $\hat{H}_{\mathrm{eff}}$ is already known from the flow-equation (\ref{eq:flow})
calculation, the problem reduces to finding the similarity transformation that
relates the two known matrices $\hat{U}_{\mathrm{strob}}^{[t]}$ and $\hat{U}_{\mathrm{eff}}(t+T,t)=
\exp[(i\hbar)^{-1}\hat{H}_{\mathrm{eff}}T]$.
We obtain $\hat{U}_{\mathrm{mic}}(\omega t)$ from the spectral decompositions of these matrices and then recover
$\hat{S}_{\mathrm{mic}}(\omega t)$ by taking the matrix logarithm with the appropriate branch choice.

\begin{figure}[h!]
    \centering
    \includegraphics[width=0.99\columnwidth]{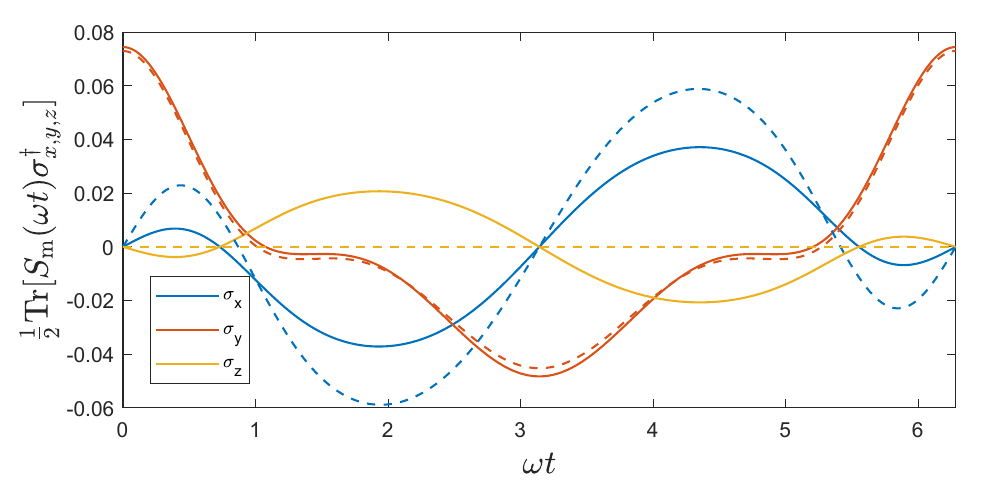}
    \caption{Comparison of the numerically obtained micromotion operator (solid lines) with the analytical expression~(\ref{eq:mic1}) (dashed lines) for various initial phases. Three projections of the micromotion vector to the basis vectors $\{\hat{\sigma}_x,\hat{\sigma}_y,\hat{\sigma}_z\}$ of $\mathfrak{su}(2)$ Lie algebra are depicted in different colors, see legend. }
    \label{fig2}
\end{figure}

The comparison between the reconstructed micromotion and the analytical first-order expression is shown in Fig.~\ref{fig2} and serves as a validation of both approaches. The $\hat{\sigma}_y$ components almost coincide, while $\hat{\sigma}_x$ and $\hat{\sigma}_z$ have some deviations. These deviations quantify the corrections not included in Eq.~(\ref{eq:mic1}).

\begin{figure*}[t!hb]
\includegraphics[width=0.95\textwidth]{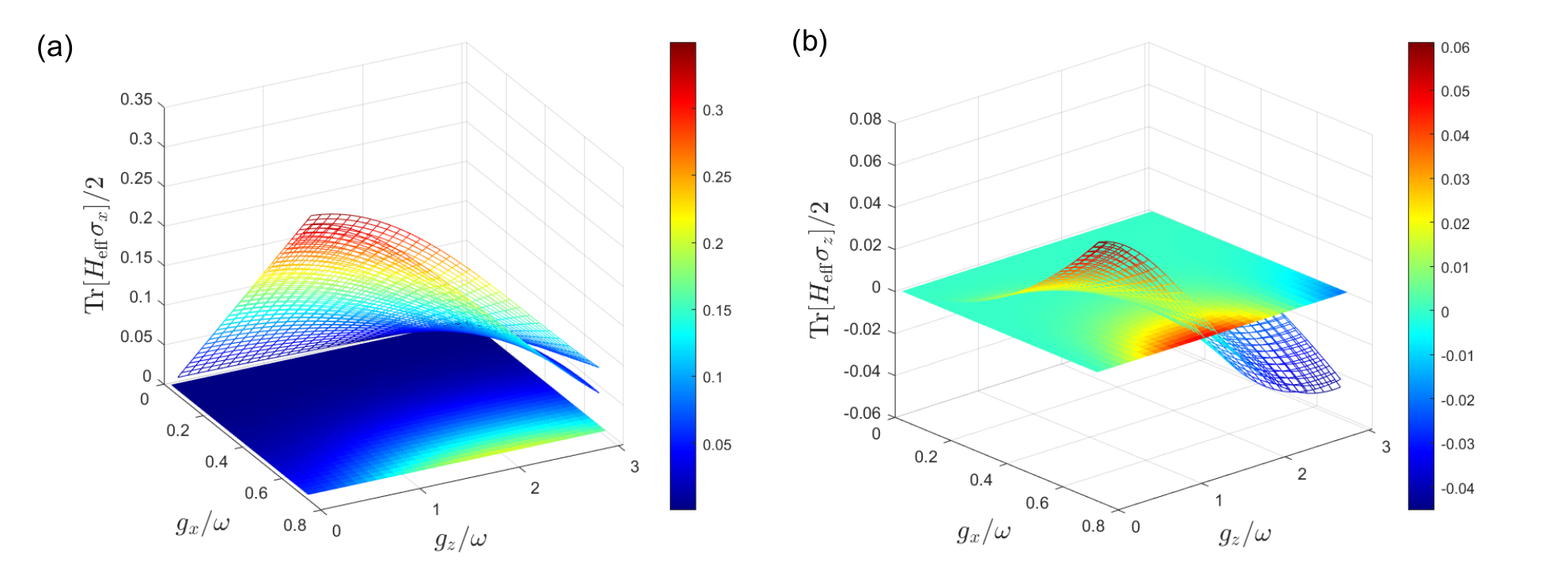}
\caption{\label{fig3} 
Comparison of the analytical and numerical effective Hamiltonians over the
two-dimensional parameter space spanned by $g_x$ and $g_z$. Panels (a) and (b)
show the coefficients multiplying $\hat{\sigma}_x$ and $\hat{\sigma}_z$,
respectively, in the decomposition (\ref{eq:Heff_components}). The flat
colored surface shows the difference between the analytical and numerical
results, multiplied by a factor of 5 for visibility.}
\end{figure*}
\subsection{Accuracy across the parameter space}
The analytical effective Hamiltonian derived from Eqs.~(\ref{eq:H0}) and~(\ref{eq:eff1_sim}) relies on the assumption that the transverse coupling remains small compared with the driving frequency. It is therefore natural to examine how the analytical expressions compare with the numerically obtained effective Hamiltonian once $g_x/\omega$ is no longer small.

To this end, we calculate the effective Hamiltonian $\hat{H}_{\mathrm{eff}}$ numerically over the
two-dimensional parameter space spanned by $g_x$ and $g_z$, and decompose it in
the Pauli basis. 
Figure~\ref{fig3} compares the analytical and numerical results for the coefficients multiplying
$\hat{\sigma}_x$ and $\hat{\sigma}_z$. The $\hat{\sigma}_y$ component is not shown, since it vanishes for both the analytical and numerical effective Hamiltonians.
The comparison confirms the expected domain of validity of the analytical expansion. For $g_x/\omega\ll1$, the analytical and numerical results agree well for both coefficients. As $g_x/\omega$ approaches unity, visible discrepancies develop as the truncated inverse-frequency expansion breaks down.
The agreement remains good over a much broader range of $g_z/\omega$, including values of order unity and larger, because the transformed-frame Hamiltonian~(\ref{eq:int_pic}) was deliberately prepared to account for large values of $g_z$.

%%%%%%%%%%%%%%%%%%%%%%%%%%%%%%%%%%%%%%%%%%%%%%%%%%%%%%%%%%%%%%%%%%%%%%%%%%%%%%%%%%%%%%%%  C O N C L U S I O N S
\section{Conclusions}

We have developed a Floquet framework for a driven polar two-level quantum system with a strong longitudinal 
coupling. The central step is a transformation to a reference frame in which the longitudinal modulation 
is absorbed nonperturbatively. In this frame, the Fourier components of the Hamiltonian are well suited 
for a van Vleck high-frequency expansion, even when $g_z/\omega$ is not small.

Within this formulation, we have derived an analytical time-independent effective Hamiltonian up to first order 
in the high-frequency expansion, together with the corresponding first-order micromotion operator. The longitudinal 
coupling enters the result through Bessel-function renormalization factors, while the transverse coupling remains 
a perturbative parameter. In the nonpolar limit, the zeroth-order Hamiltonian recovers the usual transverse 
coupling of the near-resonant Rabi model, and the first-order diagonal correction reduces to the Bloch--Siegert shift.

As a second main result, we have constructed the effective Hamiltonian numerically using flow equations. 
This approach provides a numerical counterpart to the analytical high-frequency expansion. We have used 
it to benchmark the analytical zeroth- and first-order Hamiltonians, to reconstruct the micromotion, 
and to compare the resulting effective dynamics with direct numerical integration of the Schr\"odinger equation.

The comparison confirms the expected domain of validity of the analytical description. The expansion remains 
accurate when the transverse coupling is small compared with the driving frequency, $g_x/\omega\ll1$, while 
deviations appear as $g_x$ becomes comparable to $\omega$. By contrast, the agreement can persist 
for longitudinal couplings of order $\omega$ or larger, reflecting the nonperturbative treatment 
of $g_z/\omega$ in the dressed-frame formulation.

The theory for the regime where both the transverse and longitudinal couplings are strong remains to be 
developed and is a prospect for future work.

\begin{acknowledgments}
This project received funding from the Research Council of Lithuania (LMTLT), Agreement No.~S-ITP-24-6. 
The authors are grateful to Domantas Burba for his enlightening discussion. KS is grateful to the support 
of the National Science Centre, Poland, grant number 2023/50/E/ST3/00451.
\end{acknowledgments}

\bibliography{references}% Produces the bibliography via BibTeX.

\end{document}